\documentstyle[aps,prl,floats,epsf,preprint]{revtex}
\tightenlines
\begin{document}
\draft
\title{Aharonov-Casher-Effect Suppression of Macroscopic Tunneling of Magnetic Flux}
\author{Jonathan R. Friedman$^{1,2}$ and D. V. Averin$^{1}$}
\address{$^1$Department of Physics and Astronomy, The State University \\
of New York at Stony Brook, Stony Brook, NY 11794-3800 \\
$^2$Department of Physics, Amherst College, P.O. Box 5000, Amherst, \\
MA 01002-5000}
\date{\today}
\maketitle

\begin{abstract}
We suggest a system in which the amplitude of macroscopic flux tunneling can
be modulated via the Aharonov-Casher effect. The system is an rf-SQUID with
the Josephson junction replaced by a Bloch transistor - two junctions
separated by a small superconducting island on which the charge can be
induced by an external gate voltage. When the Josephson coupling energies of
the junctions are equal and the induced charge is $q=e$, destructive
interference between tunneling paths brings the flux tunneling rate to zero.
The device may also be useful as a qubit for quantum computation.
\end{abstract}


\smallskip

Geometric-phase effects are ubiquitous in physics \cite{ref1}. The best
known of these is the Aharonov-Bohm effect \cite{ref2} in which charges
moving in a field-free region surrounding a magnetic flux nevertheless pick
up a phase that leads to interference. This effect long predates the general
formulation of the geometric phase by Berry \cite{ref3}. A related effect
was proposed by Aharonov and Casher \cite{ref4}, who showed that a magnetic
moment that moves around a line charge acquires a phase proportional to the
linear charge density. Extending this analysis to superconductors, Reznik
and Aharonov \cite{ref5} showed that magnetic vortices (or fluxons) moving
around an external charge $q$ could acquire an Aharonov-Casher phase. Other
theoretical work has concerned the appearance of this effect in
Josephson-junction arrays and devices \cite{ref6,ref7,ref9,ref10,ref12,ref13}%
. Experimental evidence for the Aharonov-Casher effect in a Josephson system
was first provided by Elion et al. \cite{ref14}. There the relevant charge
around which the vortices move is that induced on a superconducting island
by an external gate voltage. Manifestations of the Aharonov-Casher effect in
Josephson-junction systems studied up to now -- $2e$-periodic modulation of
an array resistance with the induced charge and quantization of the voltage
across the array -- are, however, not specific to this effect. Qualitatively
similar phenomena can be produced simply by the Coulomb-blockade effect due
to the quantization of charge on the superconducting islands of the devices 
\cite{ref15}. In this Letter, we suggest a system, where the Aharonov-Casher
effect manifests itself in a way that can be directly distinguished from
Coulomb-blockade oscillations.

\begin{figure}[htb]
\epsfxsize=4.3in \leavevmode
\centering
\epsfbox{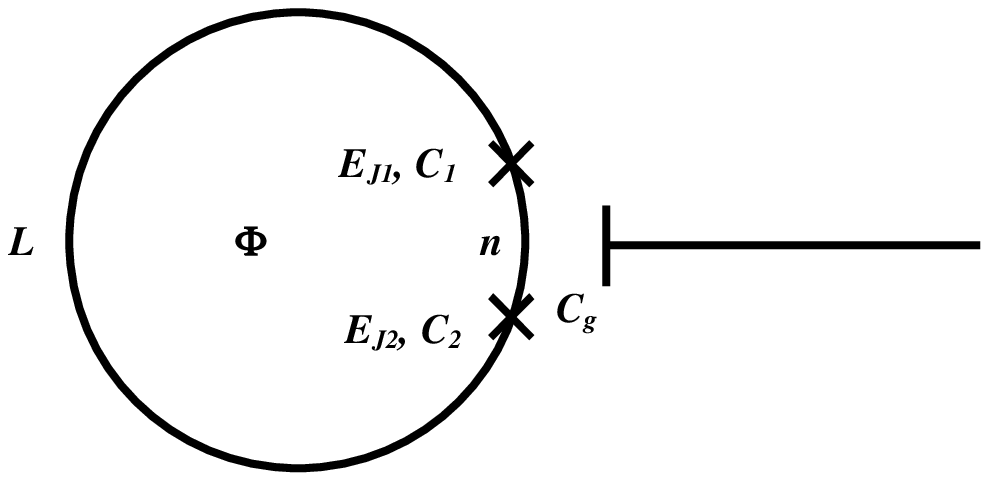} 
\caption{Schematic of the proposed device, an rf-SQUID with the
single Josephson junction replaced by a Bloch transistor. 
The charge on the superconducting island of the transistor can be 
induced by a gate voltage. 
\label{squid} }

\end{figure}

The proposed system (Fig.\ \ref{squid}) is a variant of a SQUID. Macroscopic
quantum tunneling of the magnetic flux in SQUID systems has been the subject
of extensive experimental investigation in the past several years \cite
{ref16,ref20}. Recent experiments have shown that a SQUID can be put into a
coherent superposition of macroscopically distinct flux \cite{ref20} and
persistent-current \cite{ref21} states. The system we consider is a SQUID
with the Josephson junction replaced by a Bloch transistor (see, e.g., \cite
{ref15}) -- two small, closely-spaced Josephson junctions, in which the
charge on the island between the junctions can be varied by an external gate
voltage $V_g$. Because of the gate-voltage dependence of the critical
current of the Bloch transistor, the flux-tunneling rate in such a SQUID
should exhibit $2e$-periodic oscillations as a function of the
gate-voltage-induced charge $q = C_g V_g$. As we show below, in addition to
these (``Coulomb-blockade'') oscillations, the flux tunneling is completely
suppressed when the induced charge is $e$. This fact does not follow from
the simple charge quantization on the middle electrode of the Bloch
transistor, but can be explained naturally in terms of the Aharonov-Casher
effect.

In a handwaving sense, flux tunneling occurs via the passage of a fluxon
through one of the junctions and interference can occur between the two
paths that encircle the induced charge $q$ on the middle electrode. When $%
q=e $, the relative phase of the two paths is $\pi$ and the interference is
destructive. If, in addition, the Josephson coupling energies $E_{J1} $and $%
E_{J2} $ of the two junctions (which control the amplitude of flux tunneling
through each individual junction) are equal, $E_{J1} = E_{J2} $ \cite{ref22}%
, the destructive interference is complete, and the total amplitude of flux
tunneling is exactly zero. More generally, the Aharonov-Casher effect in
this system makes it possible to control the phase of the flux-tunneling
amplitude, an effect that can simplify design of Josephson-junction qubits
for quantum computing \cite{ref20,ref21,ref1a,ref4a,ref2a,ref3a}.

Quantitatively, we consider a SQUID of inductance $L$ (Fig.\ \ref{squid})
that is assumed to be not too large, so that the SQUID has only two
metastable flux states \cite{ref23}. The SQUID has two junctions with
Josephson phase differences $\phi _1 $ and $\phi _2 $ across them. A gate
electrode couples with capacitance $C_g $ to the island between the
junctions and an external flux $\Phi _x $ is applied to the SQUID loop. The
Hamiltonian for this system is

\begin{equation}
H = \frac{{Q^2}}{{2C }} + \frac{{(2en - q)^2 }}{{2C_\Sigma }} +\frac{{(\Phi
- \Phi _x )^2 }}{{2L}} - E_{J1} \cos \phi _1 - E_{J2} \cos \phi _2 \, ,
\label{Ham}
\end{equation}
where $n$ is the number of Cooper pairs charging the island, $C_{\Sigma}$ is
its total capacitance relative to all other electrodes. The flux $\Phi $ in
the SQUID is related to the phase differences across the junctions by $2\pi
\Phi /\Phi _0 = \phi _1 + \phi _2 $, where $\Phi _0 = h/2e$ is the flux
quantum. $\Phi$ is conjugate to the charge $Q$ on the capacitance $C$
between the ends of the SQUID loop: $[\Phi,Q]=i\hbar$. (Because of the
stray-capacitance contribution to $C$, it is not related directly to the
island capacitance $C_{\Sigma}$.) Similarly, the phase $\theta = (\phi _1 -
\phi _2 )/2$ is conjugate to $n$: $[\theta ,n]=i$. Thus, $\Phi $ and $\theta 
$ are the ``coordinates'' of the SQUID and $Q$ and $n$ are the "momenta".

\begin{figure}[htb]
\epsfxsize=4.3in \leavevmode
\centering
\epsfbox{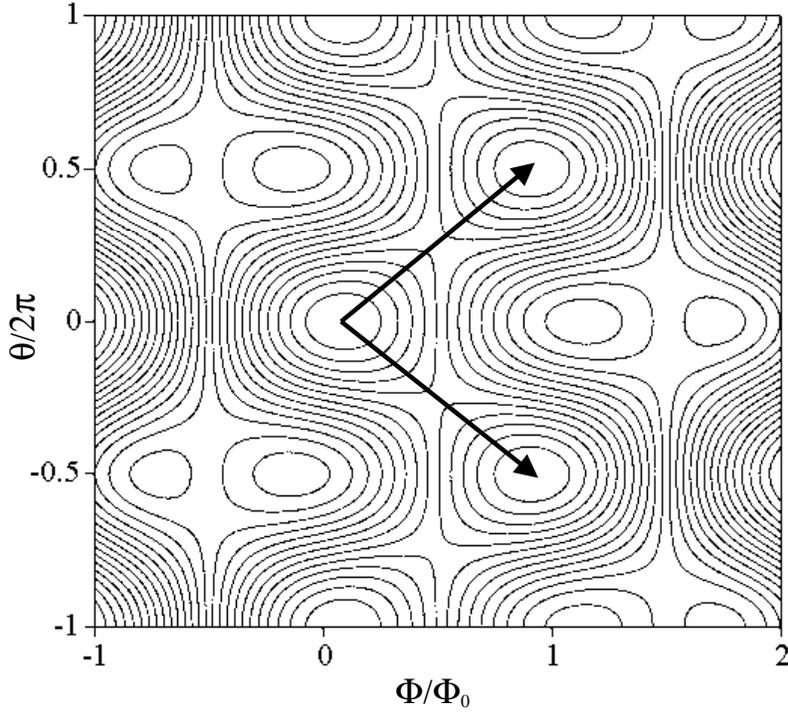} 
\caption{Two-dimensional potential for the system described by 
Eq.\ (\ref{Ham}) with $E_{J1}  = E_{J2}  = E_J $
and $E_J = 0.507\, \Phi _0^2 /2L $.
When $q=e$, destructive interference between
the two paths shown leads to suppression of flux tunneling.  
\label{Pot} }

\end{figure}

One can then view the first two terms in the Hamiltonian as representing the
kinetic energy of the system and the remaining terms as the potential
energy, shown in Fig.\ \ref{Pot} for the case $\Phi _x = \Phi _0 /2$ and $%
E_{J1}=E_{J2}$. The potential has minima in a checkerboard-like pattern with
minima on one side of $\Phi = \Phi _0 /2$ shifted by $\pi $ in the $\theta $
direction relative to minima on the other side. The minima are not separated
by integer multiples of $\Phi _0 $ in the $\Phi $ direction because of the
inductive term in Eq.\ (\ref{Ham}). On the other hand, the potential is
strictly periodic in the $\theta $ direction and, consequently, neighboring
minima are always separated by $\Delta \theta = \pm \pi $. In fact, since
the number $n$ of Cooper pairs on the island is integer, one can impose
periodic boundary conditions on $\theta $, wrapping all points with $\theta
= \pi $ to $\theta = - \pi $.

We will analyze tunneling between different flux states in two limits, one
in which the charging energy for the island $E_C = (2e)^2/2C_{\Sigma}$ is
much smaller than $E_J $ and the other in which $E_C \gg E_J $. First, we
will consider the former case, where $\theta $ is a good quantum number and
one can employ a tight-binding picture in which the system is localized in
one of the minima of Fig.\ \ref{Pot}. The tunnel splitting can be found
using an instanton technique in which the imaginary-time action $S_I $ is
evaluated along the least-action paths in the inverted potential \cite{ref24}%
. The zero-temperature tunnel splitting is then given by

\begin{equation}
\Delta = \sum\limits_j {\omega _j e^{ - S_I^j /\hbar } } \, ,  \label{delta}
\end{equation}
where the sum is over all least-action paths between two potential minima,
and $\omega _j $ and $S_I^j $ are the attempt frequency and imaginary-time
action, respectively, for the jth path. The action is calculated by
evaluating

\begin{equation}
S_I^j = \int\limits_{ - \infty }^\infty {{\cal L}_E (\tau )d\tau }
\label{inst}
\end{equation}
along the jth path, where ${\cal L}_E (\tau ) \equiv - {\cal L}(t \to -
i\tau )$ is the imaginary-time (Euclidean) Lagrangian obtained from the
real-time Lagrangian ${\cal L}$, which in turn is obtained from the
Hamiltonian by the usual transformation: ${\cal L} = \dot \Phi Q + ({%
\textstyle {\frac{{\Phi _0 } }{{2\pi }}}}\dot \theta) (2en) - H$. For $%
E_{J1} = E_{J2} = E_J $ and using Hamilton's equations to obtain $\dot \Phi
= Q /C $ and ${\textstyle {\frac{{\Phi _0 } }{{2\pi }}}}\dot \theta = (2en -
q)/C_\Sigma $ the imaginary-time Lagrangian is found to be

\begin{equation}
{\cal L}_E (\tau ) = \frac{{C} }{2}\dot \Phi ^2 + \frac{{C_\Sigma }}{2}%
\left( {\frac{{\Phi _0 \dot \theta }}{{2\pi }}} \right)^2 + \frac{{(\Phi -
\Phi _x )^2 }}{{2L}} - 2E_J \cos (\pi \Phi /\Phi _0 )\cos \theta - iq\left( {%
\frac{{\Phi _0 \dot \theta }}{{2\pi }}} \right),  \label{Lagrangian}
\end{equation}
where the dot represents differentiation with respect to $\tau $. The last
term in Eq.\ (\ref{Lagrangian}), being a total time derivative, can have no
effect on the classical dynamics of the system. Yet, it has profound effects
on tunneling, giving rise to the Aharonov-Casher phase and the resulting
interference effect.

To see this, let us consider the two tunneling paths illustrated
schematically by the arrows in Fig. \ref{Pot}. The endpoints of the two
paths are equivalent, since they merely differ in $\theta $ by $2\pi $.
Every term in Eq.\ (\ref{Lagrangian}) except the last is symmetric under the
reflection $\theta \to - \theta $. Hence, the actions of the two paths
differ only because of the last term. The imaginary-time action for each
path can then be separated into two parts: $S_I^j = \tilde S_I + S_{geo}^j $%
, where $\tilde S_I $ is the action obtained by using all but the last term
in Eq.\ (\ref{Lagrangian}) and is the same for both paths. The
geometric-phase action for each path is given by 
\begin{equation}
S_{geo}^{1,2} = - iq\left( {\frac{{\Phi _0 }}{{2\pi }}} \right)\int%
\limits_{1,2} {\dot \theta d\tau } = \mp iq\left( {\frac{{\Phi _0 }}{{2\pi }}%
} \right)\pi = \mp i\pi (q/2e)\hbar .  \label{Sgeo}
\end{equation}
From Eq.\ (\ref{delta}), the tunnel splitting is given by 
\begin{equation}
\Delta = 2\Delta _0 \cos (q\pi /2e),  \label{delta_cos}
\end{equation}
where $\Delta _0 = \omega _0 e^{ - \tilde S_I /\hbar } $ is the tunnel
splitting associated with one path. The explicit forms of parameters $\omega
_0 $ and $\tilde S_I $ are not relevant for the present calculation; they
will be presented in subsequent work.

The upshot of Eq.\ (\ref{delta_cos}) is that the tunnel splitting for flux
tunneling can be modulated by the application of a gate voltage. When the
induced charge is $e$ (half a Cooper pair), the two tunneling paths
interfere completely destructively and flux tunneling is wholly suppressed.
Furthermore, when $e < q\bmod (4e) \le 3e$ the tunnel splitting becomes
negative. This means that the ground and excited states interchange roles:
if the ground (excited) state can be approximated by $(\left| \phi_0
\right\rangle + ( - )\left| \phi_1 \right\rangle)/\sqrt{2} $ when $- e \le
q\bmod (4e) < e$, where $\left| \phi_0 \right\rangle $ and $\left| \phi_1
\right\rangle $ are the distinct fluxoid states connected by the tunneling
paths in Fig. \ref{Pot}, then it becomes $(\left| \phi_0 \right\rangle - ( +
)\left| \phi_1 \right\rangle)/\sqrt{2} $ for $e < \left| q \right|\bmod (4e)
\le 3e$.

In the limit of large ``internal'' charging energy, $E_C\gg E_J$, the
physics of suppression of flux tunneling for $q\simeq e$ remains the same as
for $E_C \ll E_J$. The quantitative form of the tunneling amplitude is,
however, quite different. The difference is due to the fact that for $E_C\gg
E_J$ the system wave function is delocalized in the $\theta$-direction and
flux trajectories with all values of $\Delta\theta$ contribute to the
tunneling rate. Since the regime of flux tunneling requires the ``external''
charging energy $E_Q = (2e)^2/2C$ to be smaller than $E_J$, all energies of
the flux dynamics are then smaller than the energies of the charge dynamics
on the central island. In this case, when $q\simeq e$, only the two charge
states of the island, $n=0$ and $n=1$, are relevant for the charge dynamics.
The Hamiltonian of the system reduces to:
\begin{equation}
H=\frac{Q^2}{2C}+\frac{(\Phi-\Phi_x)^2}{2L}-E_+ \cos (\pi \Phi/\Phi_0)
\sigma_z + \frac{e(q-e)}{C_\Sigma} \sigma_x - E_- \sin (\pi \Phi/\Phi_0)
\sigma_y\, ,  \label{a1}
\end{equation}
where $E_{\pm}\equiv (E_{J1}\pm E_{J2})/2$ and the Pauli matrices
$\sigma_i$ act on the basis of $(|0\rangle \pm |1\rangle)/\sqrt{2}$, 
the symmetric and antisymmetric superpositions of the
charge states.

The potential $U(\Phi)$ for the evolution of the flux $\Phi$ can be seen to
have two branches, depending on the charge-space state of the system: For $%
\Phi<\Phi_0/2$, the symmetric charge state $(|0\rangle + |1\rangle)/\sqrt{2}$
provides the lower branch of the potential, whereas for $\Phi>\Phi_0/2$, the
sign of $\cos (\pi \Phi/\Phi_0)$ changes and the lower branch of the
potential is given by the antisymmetric state $(|0\rangle -|1\rangle)/\sqrt{2%
}$. This means that to avoid suppression of flux tunneling from a state
localized in the region $\Phi<\Phi_0/2$ into the state with $\Phi> \Phi_0/2$%
, the system should make a transition between the symmetric and
antisymmetric charge states in the course of flux tunneling. From the wave
function of these states in the $\theta$-representation: $|\psi_+| \propto
\cos(\theta/2)$ and $|\psi_-| \propto \sin (\theta/2)$, one can see that
such a transition corresponds to the shift in $\theta$ by $\pm \pi$,
analogous to the similar shift in the limit of large $E_J$.

The Hamiltonian (\ref{a1}) shows that the coupling between the two branches
of the potential is provided by the difference $E_{-}$ of the Josephson
coupling energies and deviations of the induced charge $q$ from $e$. In the
absence of coupling, transitions between the two potential branches and, as
a result, flux tunneling, are suppressed. When the coupling is weak (i.e., $%
E_{-}$ and $q-e$ are small), transitions between the potential branches
takes place in the vicinity of the degeneracy point $\Phi =\Phi _{0}/2$ and
can be described in terms of Landau-Zener tunneling. The only difference
from the standard situation of Landau-Zener tunneling is that the transition
is not driven by a classical external force but by the flux motion under the
tunnel barrier and therefore can be described as occurring in imaginary
time. The amplitude of such an imaginary-time Landau-Zener transition has
been found in the context of the dynamics of Andreev levels in
superconducting quantum point contacts \cite{ref25}. Adapted for the present
problem, the expression for the transition amplitude is: 
\begin{equation}
w=\frac{1}{\Gamma (\lambda )}\left( \frac{2\pi }{\lambda }\right)
^{1/2}\left( \frac{\lambda }{e}\right) ^{\lambda }e^{i\phi
}\,,\;\;\;\;\lambda \equiv \frac{\lbrack e(q-e)/C_{\Sigma }]^{2}+E_{-}^{2}}{%
2E_{+}[E_{Q}E]^{1/2}}\,.  \label{a2}
\end{equation}
Here $\Gamma $ denotes the Gamma function and $E$ is the absolute value of
the energy of the quasistationary flux state at $\Phi <\Phi _{0}/2$ measured
relative to the effective top of the potential barrier formed by the
crossing at $\Phi =\Phi _{0}/2$ of the two branches of the potential. The
phase $\phi $ of the tunneling amplitude coincides with the phase of the
coupling between the potential branches in the Hamiltonian (\ref{a1}) and is
given by 
\begin{equation}
\tan \phi =\frac{E_{-}C_{\Sigma }}{e(q-e)}\,.  \label{a3}
\end{equation}

\begin{figure}[htb]
\setlength{\unitlength}{1.0in}
\begin{picture}(5.0,4.4) 
\put(0.6,.3){\epsfxsize=4.3in\epsfysize=3.4in\epsfbox{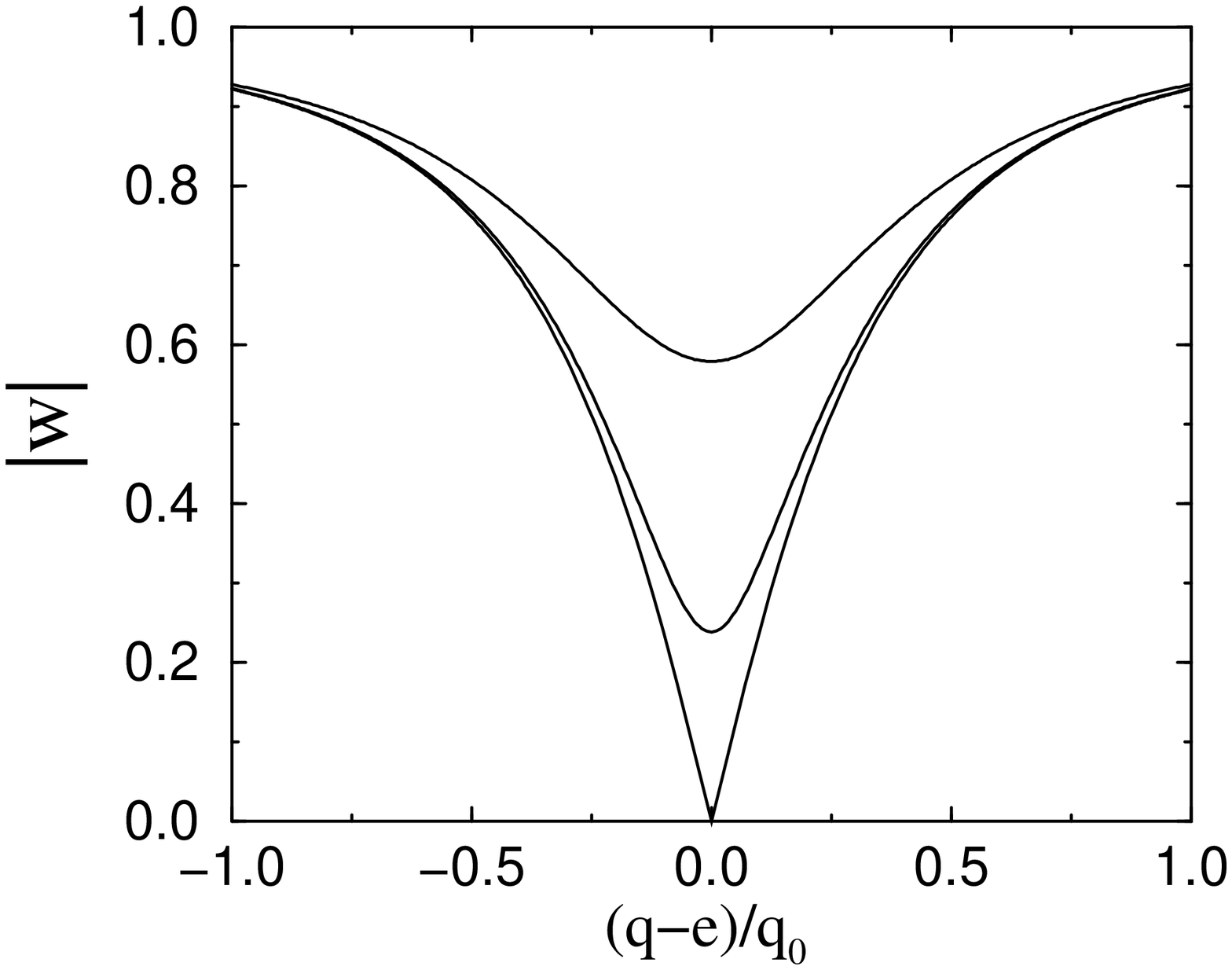}}
\end{picture}
\caption{Absolute value of the transition amplitude between 
the two branches of the flux potential, which is proportional to the 
amplitude of flux tunneling, as a function of the induced 
charge $q$ at $q\simeq e$. Note that the scale $q_0\equiv 
C_\Sigma (2E_+)^{1/2}(E_Q E)^{1/4}/e$ of variations of $q$ is much smaller 
than $e$. From bottom to top, the curves correspond to increasing 
difference between the Josephson energies of the two junctions: 
$E_-/(2E_+)^{1/2} (E_Q E)^{1/4} = 0.0\, , 0.1\, ,0.3$. 
\label{f:a} }

\end{figure}

Since the overall amplitude of flux tunneling is proportional to $w$ and
since $w(\lambda) \simeq (2\pi\lambda)^{1/2}$ for $\lambda \ll 1$, Eq.\ \ (%
\ref{a2}) shows that, when $E_-$ is small, flux tunneling is suppressed at $%
q\simeq e$. Equation (\ref{a3}) also shows that, similarly to the limit of
large $E_J$, the sign of the tunneling amplitude at $E_- \rightarrow 0$
changes abruptly when $q$ moves through the point $q=e$. The absolute value
of amplitude $w$ (\ref{a2}) is shown in Fig.\ \ref{f:a} as a function of
charge $q$ for several values of $E_-$. The curves in Fig.\ \ref{f:a}
indicate that the suppression of the flux tunneling amplitude at $q\simeq e$
remains pronounced for quite large degree of asymmetry of the Josephson
coupling energies of the two junctions.

The suppression of tunneling occurs for arbitrary $E_J/E_C$, as can be seen
from the following argument. The ground-state degeneracy responsible for the
suppression of tunneling is due to two symmetries of Hamiltonian (\ref{Ham})
that occur for $E_{J1}=E_{J2}$ when $\Phi_x=\Phi_0/2$ and $q=e$. The
Josephson coupling terms in the Hamiltonian can be written as $-2E_J \cos
(\pi\Phi/\Phi_0 ) \cos \theta=-E_J \cos (\pi\Phi/\Phi_0 ) \sum\limits_n ({%
\left| n \right\rangle \left\langle {n + 1} \right|} + \left| {n + 1}
\right\rangle \left\langle n \right|)$. Then the Hamiltonian is symmetric
with respect to the following two transformations: (I) $\left| n
\right\rangle \to ( - 1)^n \left| n \right\rangle\, , \Phi\rightarrow
\Phi_0-\Phi$, which requires that $\Phi_x=\Phi_0/2$, and (II) $n \to 1 - n$,
which requires that $q=e$. We now show that one cannot construct an energy
eigenfunction that is simultaneously an eigenfunction of both of these
transformations and, therefore, that each energy eigenfunction must be
degenerate with another one with which it is related by the application of
both transformations. Let us write the energy eigenfunction in $\Phi,n$
space: $\left| {\Psi \left( {n,\Phi } \right)} \right\rangle = \sum\limits_n 
{\psi _n \left( \Phi \right)} \left| n \right\rangle $. Transformation (I)
gives $\psi _n \left( \Phi \right) = \left( {\ - 1} \right)^n \psi _n \left( 
{\Phi _0 - \Phi } \right)$, i.e. $\psi _n \left( \Phi \right)$ is an odd
(even) function of $\Phi - \Phi _0 /2$ if $n$ is odd (even). Transformation
(II) requires that $\psi _n \left( \Phi \right) = \psi _{1 - n} \left( \Phi
\right)$, but this is impossible because it equates an even function with an
odd one. Thus, at $\Phi_x=\Phi_0/2$ and $q=e$, the eigenfunctions of
Hamiltonian (\ref{Ham}) are all two-fold degenerate for arbitrary $E_J/E_C$
ratio.

Our results are a variant of the Aharonov-Casher effect applied to
superconductors, like that studied by Reznik and Aharonov \cite{ref5}. It is
important to note that in our calculation, the relevant charge is not $n$,
which may not even be a good quantum number, but $q$, the charge induced by
the gate voltage. Another interesting feature of our results is that the
amount of flux that tunnels need not be a multiple of $\Phi _0 $, as noted
above and illustrated in Fig.\ \ref{Pot}. Thus, unlike the case studied in 
\cite{ref5} where fluxons of flux $\Phi _0 $ move around a localized,
external charge, the modulation of flux tunneling in the SQUID does not
require that the flux $\Phi$ tunnel between quantized values or that the
charge on the island be localized.

We wish to thank J. M\"{a}nnik, E. Chudnovsky and A. Guillaume for useful
discussions. This work was supported in part by the NSA and ARDA under ARO contract 
DAAD199910341.

\begin{references}
\bibitem{ref1}  {A. Shapere and F. Wilczek, Geometric Phases in Physics
(World Scientific, Singapore, 1989).}

\bibitem{ref2}  {Y. Aharonov and D. Bohm, Physical Review {\bf 115}, 485
(1959).}

\bibitem{ref3}  {M. V. Berry, Proc. R. Soc. London Ser. A-Math. Phys. Eng.
Sci. {\bf 392}, 45 (1984).}

\bibitem{ref4}  {Y. Aharonov and A. Casher, Phys. Rev. Lett. {\bf 53}, 319
(1984).}

\bibitem{ref5}  {B. Reznik and Y. Aharonov, Phys. Rev. D {\bf 40}, 4178
(1989).}

\bibitem{ref6}  {B. J. van Wees, Phys. Rev. Lett. {\bf 65}, 255 (1990).}

\bibitem{ref7}  {T. P. Orlando and K. A. Delin, Phys. Rev. B {\bf 43}, 8717
(1991).}

\bibitem{ref9}  {J. X. Zhu, Z. D. Wang and Q. W. Shi, J. Phys. A-Math. Gen. 
{\bf 27}, L875 (1994).}

\bibitem{ref10}  {R. Fazio, A. van Otterlo and A. Tagliacozzo, Europhys.
Lett. {\bf 36}, 135 (1996).}

\bibitem{ref12}  {E. Simanek, Phys. Rev. B {\bf 55}, 2772 (1997).}

\bibitem{ref13}  {A. Vourdas, Europhys. Lett. {\bf 48}, 201 (1999).}

\bibitem{ref14}  {W. J. Elion, J. J. Wachters, L. L. Sohn and J. E. Mooij,
Phys. Rev. Lett. {\bf 71}, 2311 (1993).}

\bibitem{ref15}  {D. V. Averin and K. K. Likharev, in: Mesoscopic Phenomena
in Solids, edited by B. L. Altshuler, P. A. Lee and R. A. Webb (Elsevier,
Amsterdam, 1991), p. 173.}

\bibitem{ref16}  {R. Rouse, S. Y. Han and J. E. Lukens, Phys. Rev. Lett. 
{\bf 75}, 1614 (1995).}

\bibitem{ref20}  {J. R. Friedman, V. Patel, W. Chen, S. K. Tolpygo and J. E.
Lukens, Nature {\bf 406}, 43 (2000).}

\bibitem{ref21}  {C. H. van der Wal, A. C. J. ter Haar, F. K. Wilhelm, R. N.
Schouten, C. Harmans, T. P. Orlando, S. Lloyd and J. E. Mooij, Science {\bf %
290}, 773 (2000).}

\bibitem{ref22}  {The Josephson energies can be made equal in practice by
replacing at least one junction by a small dc-SQUID, as has been done in
recent experiments \cite{ref16,ref20}, where the effective Josephson energy
can be tuned via the application of an external flux. Our analysis also
assumes that fluctuations in $Q$ do not couple to the island, a condition 
that is satified, e.g., when the two junctions have similar capacitances.}

\bibitem{ref1a}  D.V. Averin, Fortschrit. der Physik {\bf 48}, 1055 (2000).

\bibitem{ref4a}  {G. Falci, R. Fazio, G. M. Palma, J. Siewert and V. Vedral,
Nature {\bf 407}, 355 (2000).}

\bibitem{ref2a}  Yu. Makhlin, G. Sch\"{o}n, and A. Shnirman, Rev. Mod. Phys. 
{\bf 73}, 357 (2001).

\bibitem{ref3a}  K.K. Likharev, private communication.

\bibitem{ref23}  {In contrast to conventional SQUIDs, the inductance $L$ of
the double-junction SQUID considered in this work does not have to be larger
than some threshold value for the SQUID to have two metastable flux states.}

\bibitem{ref24}  {R. Rajaraman, Solitons and Instantons: An Introduction to
Solitons and Instantons in Quantum Field Theory (Elsevier Science,
Amsterdam, 1989).}

\bibitem{ref25}  {D. V. Averin, Phys. Rev. Lett. {\bf 82}, 3685 (1999);
Superlattices and Microstructures {\bf 25}, 891 (1999).}
\end{references}
\end{document}